\documentclass[12pt]{article}
\usepackage{amsfonts,amsbsy}

\begin{document}

\title{\textbf{\large The charm contribution into the proton structure
    function in DIS at the HERA \(\boldsymbol {ep} \)-collider}} 

\date{} 
\author{
  A.V.Berezhnoy, A.K.Likhoded}

\maketitle
\begin{abstract}
  In the framework of perturbative QCD and a model for the production
  of charmed hadrons the structure function \( F^{c}_{2} \) is
  calculated and compared with the experimental data of H1 and ZEUS
  Collaborations.  We show that the spectator mechanism of the \(
  D^{*} \)-meson production independent of a hadronic remnant is
  valid at \( p_{T}>10 \) GeV, only. We find that the evolution of \(
  F^{c}_{2}(x,Q^2) \) versus the virtuality $Q^2$ can be neglected in
  the kinematic region of HERA.
\end{abstract}

\section{Introduction}

As well known, the DIS data in a broad kinematic region can be
precisely described by a set of universal partonic distributions
obeyed equations of the DGLAP-evolution \cite{DGLAP} at large
virtualities \( Q^{2} \).  Some difficulties appear when one tries to
consider the heavy quark contribution into the function $F_2^c$ in the
DGALP-technique. The problem is caused by an accurate account of the
kinematic region \( Q^{2}\sim 4m_{q}^{2} \), where, on one hand, we
deal with high virtualities providing the applicability of pQCD, and,
on the other hand, the mass effects should be appropriately treated
and the heavy quarks cannot be considered as massless. Some attempts
to take into account the heavy quarks contribution near the threshold
and to match this contribution with the general DGLAP equations at
large \( Q^{2}\gg 4m_{q}^{2} \) are presented in
\cite{Ryskin,matching}. In this work we consider \( c \)-quark
production by making use of the notations in \cite{Ryskin}, where
the attempt to match two different approaches for large \( Q^{2} \)
and \( Q^{2}\sim 4m_{c}^{2} \) was done.

The \( c \)-quark contribution into the structure function in the deep
inelastic scattering can be obtained by convoluting the 
partonic distribution with the coefficient functions:
\begin{eqnarray}
\frac{1}{x}F_{2}^{c}(x,Q^{2}) & = & C_{g}(Q^{2},\mu ^{2})\otimes 
f_{g}(\mu ^{2})[x]+\nonumber \\
 &  & +C_{c}(Q^{2},\mu ^{2})\otimes f_{c}(\mu ^{2})[x]+\nonumber \label{F2c} \\
 &  & +C_{q}(Q^{2},\mu ^{2})\otimes f_{q}(\mu ^{2})[x],\label{f2c} 
\end{eqnarray}
where \( f_{i} \) denote the partonic densities in the proton, the symbol \(
\otimes \) corresponds to the convolution over the variable \( x \): 
\[ a\otimes b[x]=\int _{x}^{1}\frac{dz}{z}a(z)b\left( \frac{x}{z}\right) .\]

At small \( \mu ^{2}<\mu _{c}^{2} \) the charm contents is \( f_{c}(x,\mu
^{2})=0 \), and the structure function \( F_{c}^{2} \) is completely described
by the photon-gluon fusion \( \gamma ^{*}g\rightarrow c\bar{c} \), that gives
\(C_{g}(Q^{2},\mu ^{2})\otimes f_{g}(\mu ^{2}) \). The second term in
(\ref{f2c}) corresponds to the process of virtual photon scattering on the \( c
\)-quark from the sea of the initial hadron. At large \(
Q^{2} \) the process \( \gamma ^{*}c\rightarrow c{g} \) becomes
dominant. The process \( \gamma ^{*}q\rightarrow c\bar{c}q \) is usually not
taken into account because its contribution is suppressed by the additional
power of \( \alpha _{s} \), however below we show that this process can play an
essential role in the region of moderate \( p_{T} \) and \( Q^{2} \).

In addition to the problem of matching the different approaches in the 
kinematic regions of \( Q^{2}\sim m_{c}^{2} \) and \( Q^{2}\gg 4m_{c}^{2} \),
another important problem is connected to the comparison of theoretical
predictions with the experimental data. So, \( F^{2}_{c} \) is reconstructed in
the ZEUS and H1 experiments \cite{F2c} over the \( D^{*\pm } \)-meson
production data obtained in two decay modes: \( D^{*}\rightarrow K2\pi \), \(
D^{*}\rightarrow K4\pi \). The \( D^{*} \)-meson production cross section is
measured in the restricted kinematic region: at ZEUS one puts \( 1<Q^{2}<600
\)~GeV\( ^{2} \), \( |\eta (D^{*})|<1.5 \), \( 1.5<p(D^{*})<15 \)~GeV, \(
0.02<y<0.7 \), while the H1 Collaboration uses the similar kinematic region: \(
1<Q^{2}<100 \)~GeV, \( |\eta (D^{*})|<1.5 \), \( 1.5<p_{T}(D^{*})<10 \)~GeV, \(
0.05<y<0.7 \). The cross section in the whole kinematic region is reconstructed
under some model assumptions of charm production. One can conclude that the
really measured quantities are the spectra of the charmed mesons in the
restricted kinematic region, but the \( c \)-quark ones. That is why one has to
define a procedure of the charmed quark hadronization in order to compare the
calculated function \( F_{2}^{c}(Q^{2},x) \) with the experiment.

A procedure commonly used is based on the factorization theorem
applied for the momentum spectrum of \( D^{*} \)-meson: 
\begin{equation}
\label{frag}
\frac{d\sigma _{D^{*}}}{dp_{T}}=\frac{d\sigma _{c}}{dk_{T}}\otimes D(z),
\end{equation}
where \( D(z) \) is the fragmentation function of \( c\rightarrow D^{*}
\), and \( z=P_{T}/k_{T} \). The form of $D(z)$ is given by some kind of 
phenomenological anz\"atze \cite{frag}. Equation (\ref{frag}) should be a 
good approximation at large \( p_{T} \) and \( k_{T} \), where the
factorization theorem is valid. This theorem cannot be applied to the whole
kinematic region since this could generate essential errors in the relation
between the \( D^{*} \)-meson and \( c \)-quark spectra. The drawback of
(\ref{frag}) is clearly seen in the H1 and ZEUS data: there is a visible excess
of events in the region of pseudorapidity distribution towards the
fragmentation of proton-beam. One can see the same features in the distribution
over the variable \( z_{D^{*}}=(Pp_{D^{*}})/(Pq) \), where \( P \), \(
p_{D^{*}} \), \( q \) are four-momenta of the proton, \( D^{*}
\)-meson and virtual photon, correspondingly. The discrepancy between
next-to-leading-order (NLO) QCD prediction under the hadronization mechanism of
(\ref{frag}) and the experimental data is more essential at large \( \eta
_{D^{*}} \) and \( z_{D^{*}} \).

Usually one tries to improve the data description by using some additional
model assumptions incorporated in Monte-Carlo codes of event generator
\cite{F2c}.

\section{The charm production model}

We will discuss the model where the attempt to take into account the
hadronization from the very beginning is done. Let us consider the all
pQCD \( O(\alpha \alpha _{s}^{3}) \)-diagrams, describing the perturbative
production of the \( c\bar{q} \)-pair with the quantum
numbers of the appropriate charm mesons.

A complete set of diagrams is presented in Fig. 1. In this chapter our
consideration is restricted by the case of \( D^{*} \)-meson production in the
photon-gluon interaction \( \gamma ^{*}g\rightarrow D^{*}+\bar{c}+q \) (see
equation (\ref{f2c})). One needs the additional parameter \( \langle O\rangle
\) to take into account the contribution by the simultaneous production of \( c
\) - and \( q \)-quarks. This parameter describes the fusion of \( c \)-quark
and its light co-mover \( \bar{q} \) into \( (c\bar{q}) \)-system, and that is
proportional to the \( D^{*} \)-meson wave function squared.  This parameter
can be extracted from the experimentally known value of the fragmentation
probability $W$ of \( c\rightarrow D^{*} \) at large transverse momenta, \(
W=0.23 \) \cite{ee}.

The \( (c\bar{q}) \)-system can be in two color states: octet or singlet.
In our previous analysis of the data on the charm photoproduction and
electroproduction \cite{BKL} we considered the production of color
octet and color singlet states independently. So, we used two fusion
parameters: \( \langle O_{(1)}\rangle \) and \( \langle O_{(8)}\rangle
\), correspondingly. The ratio of these values is close to unit. At
small \( p_{T} \) the color-singlet contribution into the cross section as well
as the color-octet one behave as \( 1/p_{T}^{6} \), and they are approximately
equal to each other. At large \( p_{T} \) both contributions behave like (\(
1/p_{T}^{4} \)), and the singlet term is dominant. The color-singlet dominance
takes place due to the color-coefficient ratio. The both contributions at large
\( p_{T} \) come to the fragmentation regime. In this regime the octet and
singlet have a similar behavior, and the production process can be
characterized by the only parameter:
\[
\langle O^{\textrm{eff}}\rangle =\langle O_{(1)}\rangle
+\frac{1}{8}\langle O_{(8)}\rangle \] .
A more detail description of the model is represented in the Appendix.

One of essential features of this model consists of obeying the factorization
theorem at large transverse momenta. In other words, at large \( p_{T} \) the
formula of (\ref{frag}) describes the process with a good approximation, and
the process looks like a fragmentation.  At small momenta the fragmentation
form of the charm production is strongly broken by additional terms \( \sim
1/p_{T}^{6} \) in contrast to the fragmentation term, which behaves like \(
1/p_{T}^{4} \) as follows from (\ref{frag}).

One can see from Fig. 2 how the factorization theorem works at large
\( p_{T} \) for both the singlet contribution and the octet one. The
contributions of complete diagram-set (Fig. 1) in comparison with the
contribution of the fragmentation diagram one (as in (\ref{frag})) are
presented in the figure for both the singlet and octet.

The calculations have been done at rather large value of \( s_{\gamma
g}=(p_{\gamma }+p_{g})^{2}=200^{2} \)GeV\( ^{2} \) in order to reach the
factorization regime (\ref{frag}). One can see from the figure that there is
the region of \( p_{T} \), where the fragmentation mechanism is not valid, and
the contribution of perturbative recombination is dominant. The presence of the
recombination contribution is due to the gauge invariance of QCD. The \( p_{T}
\)-dependence of this contribution has the additional factor \( 1/p_{T}^{2} \),
that is why the recombination vanishes at large \( p_{T} \), and the \( D^{*}
\)-meson production is described by the factorization formula of (\ref{frag}).

\section{The \({D}^{*}\)-meson electroproduction}

As was mentioned above, the model under consideration has been used to describe
the charm photoproduction data of ZEUS Collaboration \cite{ZEUS}. The
fragmentation probability \( W(c\rightarrow D^{*}) \) was extracted from the
experimental data on the \( D^{*} \)-meson production in \( e^{+}e^{-}
\)-annihilation \cite{ee}. The value of factorization scale for the operators
\( \langle O_{(1,8)}\rangle \) has been fixed at the \( D^{*} \)-meson mass: \(
\mu _{F}=m_{D^{*}} \). The light quark mass equals \( m_{\bar{q}}=0.3 \)~GeV,
and the \( c \) -quark mass equals \( m_{c}=1.5 \)~GeV.

The factorization probability \( W(c\rightarrow D^{*})=0.23 \), while \(
\langle O^{\textrm{eff}}\rangle =0.25 \)~GeV\( ^{3} \). The ratio between the
octet and singlet operators has been chosen equal to \( \langle O_{(8)}\rangle
/\langle O_{(1)}\rangle =1.3 \). It is worth to mention that taking into
account the color-octet contribution essentially improves the description of
the experimental data, especially, in the description of the pseudorapidity
distribution. In this distribution the octet enforces the production into the
forward semisphere (toward the direction of the initial proton) and improves
the data description in comparison with the NLO prediction \cite{BKL}.

The calculations of cross section for the \( D^{*} \)-meson electroproduction
have been made by practically the same manner as for the photoproduction. All
we need is to replace the density-matrix of real photon by the density-matrix
of virtual photon. So, after averaging over the photon and electron
polarizations the \( D^{*} \)-meson elecroproduction amplitude squared has the
following from:
\begin{equation}
\label{A}
|A|^{2}=\sum
_{ij}\frac{k_{1}^{i}k_{2}^{j}+k_{1}^{j}k_{2}^{i}-\frac{Q^{2}}{2}g^{ij}}{Q^{4}}M
_{i}M_{j}^{*},
\end{equation}
where \( |A|^{2} \) is the \( D^{*} \)-meson electroproduction amplitude
squared, \( M_{i} \) is the \( D^{*} \)-meson photoproduction amplitude (\( i
\) is the Lorentz index of the photon polarization), \( k_{1} \)and \( k_{2} \)
are the initial and final positron momenta, and \( Q^{2}=-(k_{1}-k_{2})^{2} \).
The calculation results are presented in Figs. 3 and 4 in comparison with the
experimental data of the H1 and ZEUS Collaborations, correspondingly. The
calculations have been done with the same cuts as in the appropriate
experiments. The singlet contribution is plotted by the dashed curve, the octet
contribution is plotted by the dotted curve; the solid curve corresponds to the
sum of these contributions.

We have used two values of the scale for the running coupling constant of QCD
in our calculations: \( \mu _{R}=\sqrt{m_{D^{*}}^{2}+Q^{2}} \) (upper curve)
and \( \mu _{R}=\sqrt{4m_{D^{*}}^{2}+Q^{2}} \) (lower curve) and the scale
value \( \mu _{F}=\sqrt{m_{D^{*}}^{2}+Q^{2}} \) for the CTEQ4 parameterization
of the structure function. As one can see from the figures the model prediction
is in agreement with the experimental data. Such the good description is
achieved due to taking into account the octet contribution. This contribution
improves the distributions over the pseudorapidity \( \eta (D^{*}) \) (as in
the photoproduction); over the variable \( z(D^{*}) \) measured in the H1
experiment, and over the variable \( x(D^{*})=|{\vec{p}}^{*}(D^{*})|/W \)
measured in the ZEUS experiment (\( {\vec{p}}^{*}(D^{*}) \) is the \( D^{*}
\)-meson 3-momentum in the c.m.s. of initial virtual photon and proton). It is
worth to mention another feature of the octet contribution. It is essential at
small \( p_{T} \) and becomes negligible at large ones, where it is suppressed
by the color-factor 1/8.

Thus, in the framework of the model under consideration we have achieved a good
description for the ZEUS charm-photoproduction data \cite{ZEUS} as well as for
the charm-electroproduction data of H1 and ZEUS Collaborations \cite{F2c}. This
circumstance allows us to suppose that the extrapolation of the experimental
data into the total kinematic region with the help of our model is rather
reliable. It means, that we can calculate the total cross section production
and, therefore, extract \( F_{2}^{c} \).

\section{The structure function \( F_{2}^{c}\)}

The charm contribution \( F_{2}^{c} \) into the structure function \(
F_{2} \) is defined by the doubly differential cross-section of the
charm-production as follows:
\begin{equation}
\label{f2csig}
\frac{d^{2}\sigma ^{c}(Q^{2},x)}{dxdQ^{2}}=\frac{2\pi \alpha ^{2}_{s}}{\alpha
Q^{4}}\left\{ \left[ 1+(1-y)^{2}\right]
F_{2}^{c}-y^{2}F_{L}^{c}(Q^{2},x)\right\} 
\end{equation}
Generally one neglects the contribution of the longitudinal component \( F_{L}
\) because of its suppression. As was mentioned in the Introduction, \(
F_{2}^{c} \) is reconstructed on the base of the experimental data for the \(
D^{*} \)-meson production in the experimentally available kinematic region. The
observed production cross section is extrapolated into the total kinematic
region in the framework of some model. Thus, it is clear that the \( F^{c}_{2}
\) value depends on a model. In the framework of our model we have an
opportunity to calculate the cross section in the total kinematic region,
determine \( F^{c}_{2} \), and compare it with the results of ZEUS and H1
Collaborations.

The experimental dependence of \( F^{c}_{2} \) on \( x \) for different \(
Q^{2} \) values is shown in Fig. 5 by dots (the H1 Collaboration). These data
were extracted by the extrapolation of the experimentally observed cross
section
into the total kinematic region on the base of NLO calculations and Monte-Carlo
programs taking into account the hadronization.

The curves in this figure correspond to our model predictions. \( F_{2}^{c} \)
has been calculated according to formula (\ref{f2csig}). The falls on the
distribution tails appear because of the phase-space borders for the given
value of the \( ep \) -interaction energy and chosen values for the quark
masses.

The ZEUS experimental data on \( F^{c}_{2} \) are shown in Fig. 6 in
comparison with the predictions of our model .

\section{The perturbative recombination}

One can see that the operator expansion of (\ref{f2c}) contains not only
the term interpreted as photon-gluon production of charm, but also
that of describing the photon-quark production of charm.
Generally one neglects this term due to the additional factor of \(
\alpha _{s} \).

The diagrams, which correspond to the photon-quark term in our approach, are
shown in Fig. 7. For such kind of diagrams the only difference between the
color-octet production cross section and the color-singlet one is due to the
overall color factor of 1/8. That is why we can restrict ourselves by
considering the singlet production, only.

The recent analysis demonstrates that at large transverse momenta \( p_{T} \)
the photon-quark production contribution is suppressed by an additional factor
\( 1/p_{T}^{2} \). At small \( p_{T} \) the suppression is absent, and the
differential cross section of the \( (\bar{c}q) \) pair production at the angle
\( \Theta =0 \) (toward the direction of initial quark \( q \)) has a large
numerical coefficient in comparison with the \( c \)-quark production in the \(
g\gamma \)-interaction \cite{Braaten}: 
\begin{equation} \frac{d\hat{\sigma }(\gamma +q\rightarrow
(\bar{c}q)+c)}{d\hat{\sigma }(\gamma +g\rightarrow \bar{c}c)}\simeq
\frac{256\pi }{81}\alpha _{s}.
\end{equation}

Thus, the smallness of \( \alpha _{s} \) in the photon-quark production cross
section is compensated by large numerical coefficient. To the same moment, the
production at the angle \( \Theta =\pi \) is suppressed by the additional power
of energy. Another circumstance is essential, too: this contribution slightly
depends on the light quark mass and does not vanish at \( m_{q}\rightarrow 0
\).

In paper \cite{BKL} the \( c\bar{q} \)-pair production cross section in the
interaction of the photon and valence quark from the initial proton has been
calculated to evaluate the \( D^{+}/D^{-} \) and \( D^{0}/\bar{D}^{0} \)
asymmetries. The predicted value of asymmetry in the kinematic regions
researched by the ZEUS and H1 Collaborations is about 2-3\%, that is in the
same order of magnitude as experimental errors. At low energies of \( \gamma p
\)-interactions, the role of the photon-quark production becomes essential and
this contribution yields the asymmetry prediction which is in a good agreement
with the experimental data \cite{Braaten}. If the production asymmetry is due
to the perturbative recombination indeed, then the asymmetry decreases with the
\( p_{T} \) increase, because the perturbative recombination has the additional
\( 1/p_{T}^{2} \) factor in comparison with the leading contributions.  On the
other hand, if the asymmetry is due to the interaction between the \( c
\)-quark and the valence quark from the initial hadron, then the interaction
between the \( c \)-quark and the light quark from the initial hadron sea
exists. Such the contribution has been calculated in the framework of our
model, and it surprisingly looks like the octet contribution in the
distribution shape as well as in the absolute value. It is worth to mention
that the quark-photon contribution does not contain an additional normalization
factor, which the octet one contains. In Fig. 7 both the octet and quark-photon
contributions into the \( D^{*} \)-meson production are presented for the
kinematic region investigated by the H1 Collaboration. One can see that the
distributions over \( \log _{10}(x) \) and \( p_{T} \) are practically the same
for the whole investigated range. One can see the only small difference in the
normalization. The \( Q^{2} \)-distributions at \( Q^{2}<10 \) GeV\( ^{2} \)
have practically the qualitatively similar behavior, too. The differences
between the distributions over \( W \), \( \eta (D^{*}) \) and \( z(D^{*}) \)
are more essential for the production mechanisms compared. However, these
distributions have qualitatively analogous behavior. The singlet \( c\bar{q}
\)-pair yield in \( \gamma g \)- and \( \gamma \bar{q} \)-interactions is
presented in Fig. 8 for the kinematic conditions of H1 experiment. It is clear
from this figure that at not very large \( Q^{2} \) the sum of the \( \gamma g
\)- and \( \gamma \bar{q} \)-contributions into the singlet describes the data
so good as the sum of \( \gamma g \)-contributions into the singlet and octet.

Therefore, the singlet \( c\bar{q} \)-pair production mechanism is enough to
describe the charm photoproduction and electroproduction data in the HERA
experiments. The photon-gluon contribution as well as the quark-photon one play
essential role for the singlet \( c\bar{q} \)-pair production at the HERA
interaction energy. The serious arguments to take into account the sea quark
contribution into the charm production was presented in paper \cite{Braaten},
where the charm production asymmetry has been successfully described in the
framework of perturbative recombination in the E687 and E691 experiments.

\section{Conclusions}

The model under consideration is based, at first, on the heavy-quark production
in the perturbative theory, and second, it uses the nonperturbative model of
quarks fusion into the hadron. This model allows us to describe the existing
data on charm photoproduction and electroproduction in the total kinematic
region.

Let us itemize the main features of the model predictions:

\begin{enumerate}
\item In the region of large transverse momenta of the \( c\bar{q}
  \)-pair our model predictions coincide with the predictions of the
  factorization model in form (\ref{frag}), or in other words, the
  momentum spectrum of \( c\bar{q} \)-pair is calculated by convolving
  the heavy quark spectrum with the fragmentation function.
\item In the region of small transverse momenta and small \( Q^{2} \)
  the main contribution into the inclusive spectrum is due to the
  recombination diagrams, that depends on \( p_{T} \) as \(
  1/p_{T}^{6} \), in contrast to the fragmentation ones, which 
  depend on \( p_{T} \) as \( 1/p_{T}^{4} \). The contributions of \(
  g\gamma \)- and \( q\gamma \)-interactions into the charm
  production are comparable.
\item It is not necessary to include the term \( C_{c}\otimes c \) of
  \( F_{2}^{c} \) (the second term in expansion(\ref{f2c})) into
  the data description at the HERA energies. According to the paper
  \cite{Ryskin} the term \( C_{c}\otimes c \) becomes essential at
  large \( Q^{2} \).
\end{enumerate}
Therefore our main conclusion is formulated as follows: the spectator
character of the \( D^{*} \)-meson production, which is independent
of the flavor of the initial hadron remnant, becomes dominant at \(
p_{T}>10 \) GeV. At small transverse momenta the essential part of the
cross section is due to the interaction with the initial hadron
remnant. The interaction with the valence quark of the remnant
explains the experimentally observed flavor asymmetry in the charm yield.

\section{Acknowledgments}
We thank V. Kiselev for useful discussions. This work was in part supported 
by the Russian foundation for basic research,
grants 01-02-99315, 01-02-16585 and 00-15-96645, the Russian Ministry on the
education, grant E00-3.3-62.

\section*{Appendix}

\subsection*{ A. The hard production of four quarks in the photon-gluon
subprocess}

As was mentioned above, in the model under consideration one
supposes that the both valence quarks, the heavy and light ones, are
produced in the hard process.

In the framework of the tree level approach to the subprocess \(
g(p_{g})+\gamma (p_{\gamma })\rightarrow c(q_{c}) + \bar{c}(q_{\bar{c}}) +
d(q_{q}) + \bar{q}(q_{\bar{q}}) \) 24 Feynman diagrams contribute (see Fig. 1;
he bold line corresponds to the \( c \)-quark, the thin line corresponds to the
\( d \)-quark). In this paper the amplitude calculation has been done by
straightforward multiplying of \( \gamma \)-matrices, spinors and polarization
vectors.

Let us introduce the following notations:
\begin{eqnarray}
  & q_{c}^{2}=q_{\bar{c}}^{2}=m_{c}^{2},\qquad
  q_{q}^{2}=q_{\bar{q}}^{2}=m_{q}^{2}, & \nonumber \\
  & q_{c\bar{c}}=-q_{c}-q_{\bar{c}},\qquad
  q_{q\bar{q}}=-q_{q}-q_{\bar{q}},\qquad p_{g\gamma }=p_{g}+p_{\gamma }, &
  \nonumber \\
  & k_{g,c\bar{c}}=p_{g}+q_{c\bar{c}},\qquad
  k_{g,q\bar{q}}=p_{g}+q_{q\bar{q}}. &
\end{eqnarray}
The three-gluon vertex has the form
\begin{eqnarray}
  \Gamma ^{\mu }(p_{a},p_{b},\epsilon _{a},\epsilon _{b}) & = & -\Bigl
  ((2p_{a}+p_{b})\cdot \epsilon _{b}\Bigr )\epsilon _{a}^{\mu
  }+{(p_{a}-p_{b})}^{\mu }\Bigl (\epsilon _{a}\cdot \epsilon _{b}\Bigr
  )\nonumber \\
  &  & +\Bigl ((2p_{b}+p_{a})\cdot \epsilon _{a}\Bigr )\epsilon _{b}^{\mu
  },\nonumber \\
  \Gamma '^{\mu }(p_{a},p_{b},\epsilon _{a},\epsilon _{b}) & = &
  \Gamma ^{\mu }(p_{a},p_{b},\epsilon _{a},\epsilon
  _{b})/(p_{a}+p_{b})^{2}.
\end{eqnarray}
The bi-quark currents are defined as follows:
\begin{eqnarray} J^{\mu
    }_{c\bar{c}}=\frac{\bar{u}(q_{c})\gamma ^{\mu
      }v(q_{\bar{c}})}{q^{2}_{c\bar{c}}},\qquad \qquad J^{\mu
    }_{q\bar{q}}=\frac{\bar{u}(q_{q})\gamma ^{\mu
      }v(q_{\bar{q}})}{q^{2}_{q\bar{q}}}.
\end{eqnarray}
Also we need to define the auxiliary spinors
\begin{eqnarray}
  \displaystyle \bar{u}_{c(q)g(\gamma )}=\bar{u}(q_{c(q)})\hat{\epsilon
  }_{g(\gamma )}\frac{(\hat{q}_{c(q)}-\hat{p}_{g(\gamma
  )}+m_{c(q)})}{(q_{c(q)}-p_{g(\gamma )})^{2}-m_{c(q)}^{2}}, &  &
  v_{\bar{c}(\bar{q})g(\gamma )}=\frac{(\hat{p}_{g(\gamma
  )}-\hat{q}_{\bar{c}(\bar{q})}+m_{\bar{c}(\bar{q})})}{(p_{g(\gamma
  )}-q_{\bar{c}(\bar{q})})^{2}-m_{\bar{c}(\bar{q})}^{2})}\hat{\epsilon
  }_{g(\gamma )}v(q_{i}),\nonumber \\
  \bar{u}_{c,q\bar{q}}=\frac{\bar{u}(q_{c})\hat{J}_{q\bar{q}}(\hat{q}_{c}-\hat{
  q}_{q\bar{q}}+m_{c})}{(q_{c}-q_{q\bar{q}})^{2}-m_{c}^{2}}, &  &
  \bar{u}_{d,c\bar{c}}=\frac{\bar{u}(q_{q})\hat{J}_{c\bar{c}}(\hat{q}_{q}-\hat{
  q}_{c\bar{c}}+m_{q})}{(q_{q}-q_{c\bar{c}})^{2}-m_{q}^{2}},\nonumber \\
  v_{\bar{c},q\bar{q}}=\frac{(\hat{q}_{q\bar{q}}-\hat{q}_{\bar{c}}+m_{c})\hat{J
  }_{q\bar{q}}v(q_{\bar{c}})}{(q_{q\bar{q}}-q_{\bar{c}})^{2}-m_{c}^{2}},
  & &
  v_{\bar{q},c\bar{c}}=\frac{(\hat{q}_{c\bar{c}}-\hat{q}_{\bar{q}}+m_{q})\hat{J
  }_{c\bar{c}}v(q_{\bar{q}})}{(q_{c\bar{c}}-q_{\bar{q}})^{2}-m_{q}^{2}},
  \label{sub}
\end{eqnarray}
where \( \epsilon _{g} \) and \( \epsilon _{\gamma } \) are the
polarization vectors of the gluon and photon, correspondingly.
The matrix element squared is summed over the following ortonormalized
states of gluon: 
\[ \epsilon '=(0,1,0,0),\, \, \, \, \epsilon ''=(0,0,1,0).\]
It is worth to mention that \( {\epsilon '}^{2}={\epsilon ''}^{2}=-1
\), and \( p\cdot \epsilon =0 \), where \( p \) is gluon momentum.

In the case of deep inelastic production the photon is off mass shell,
and one need to replace the matrix \( \hat{\epsilon }_{2} \) in
(\ref{sub}) by the matrix \( \gamma ^{i} \). Thus the matrix
element will have a free Lorentz index for convolving with the
photon density in accordance with formula (\ref{A}).

We use the following index definitions:

the upper indices \( j_{g} \) designate the color state of gluon;

the low index \( i_{c} \) designates the color state of \( c \)-quark;

the low index \( i_{\bar{c}} \) designates the color state of \(
\bar{c} \)-quark;

the low index \( i_{q} \) designates the color state of \( d \)-quark;

the low index \( i_{\bar{q}} \) designates the color state of \(
\bar{q} \)-quark.

\noindent
The contributions of the Feynman diagrams into the total amplitude
can be written as follows (\(e_q\) and \( e_c\) are electric charges of 
the \( q\)- and \(c\)-quarks correspondingly; 
the color coefficients are put into the braces):

\newcommand{\F}[6]{f^{{#1_#2}{#3_#4}{#5_#6}}}
\newcommand{\T}[6]{t^{#1_#2}_{{#3_#4}{#5_#6}}}

\begin{eqnarray}
 &  & T_{1}=e_c\cdot \bar{u}(q_{q})\hat{\Gamma }'(p_{g},q_{c\bar{c}},\epsilon
 _{g},J_{c\bar{c}})v_{\bar{q}\gamma }\cdot \{i\F {n}{1}{n}{2}{j}{g}\T
 {n}{1}{i}{c}{i}{{\bar{c}}}\T {n}{2}{i}{d}{i}{{\bar{q}}}\},\\
 &  & T_{2}=e_c\cdot \bar{u}_{q\gamma }\hat{\Gamma }'(p_{g},q_{c\bar{c}},\epsilon_{g},J_{c\bar{c}})v(q_{\bar{q}})\cdot \{i\F {n}{1}{n}{2}{j}{g}\T
 {n}{1}{i}{c}{i}{{\bar{c}}}\T {n}{2}{i}{d}{i}{{\bar{q}}}\},\\
 &  & T_{3}=e_q\cdot \bar{u}(q_{c})\hat{\Gamma }'(p_{g},q_{q\bar{q}},\epsilon
 _{g},J_{q\bar{q}})v_{\bar{c}\gamma }\cdot \{i\F {n}{1}{n}{2}{j}{g}\T
 {n}{1}{i}{d}{i}{{\bar{q}}}\T {n}{2}{i}{c}{i}{{\bar{c}}}\},\\
 &  & T_{4}=e_q\cdot \bar{u}_{c\gamma }\hat{\Gamma }'(p_{g},q_{q\bar{q}},\epsilon
 _{g},J_{q\bar{q}})v(q_{\bar{c}})\cdot \{i\F {n}{1}{n}{2}{j}{g}\T
 {n}{1}{i}{d}{i}{{\bar{q}}}\T {n}{2}{i}{c}{i}{{\bar{c}}}\},\\
 &  & T_{5}=e_c\cdot \bar{u}_{cg}\gamma ^{\alpha }v(q_{\bar{c}})\bar{u}(q_{q})\gamma
 _{\alpha }v_{\bar{q}\gamma }/k_{g,c\bar{c}}^{2}\cdot \{\T
 {j}{g}{i}{c}{l}{{}}\T {n}{{}}{l}{{}}{i}{{\bar{c}}}\T
 {n}{{}}{i}{q}{i}{{\bar{q}}}\},\\
 &  & T_{6}=e_c \cdot 
 \bar{u}(q_{c})\gamma ^{\alpha }v_{\bar{c}g}\bar{u}(q_{q})\gamma_{\alpha }
 v_{\bar{q}\gamma }/k_{g,c\bar{c}}^{2}\cdot \{\T{n}{{}}{i}{c}{l}{{}}
 \T {j}{g}{l}{{}}{i}{{\bar{c}}}
 \T{n}{{}}{i}{q}{i}{{\bar{q}}}\},\\
 &  & T_{7}=e_c\cdot \bar{u}_{cg}\gamma ^{\alpha }v(q_{\bar{c}})\bar{u}_{q\gamma }\gamma
 _{\alpha }v(q_{\bar{q}})/k_{g,c\bar{c}}^{2}\cdot \{\T {j}{g}{i}{c}{l}{{}}\T
 {n}{{}}{l}{{}}{i}{{\bar{c}}}\T {n}{{}}{i}{q}{i}{{\bar{q}}}\},\\
 &  & T_{8}=e_c\cdot \bar{u}(q_{c})\gamma ^{\alpha }v_{\bar{c}g}\bar{u}_{q\gamma }\gamma
 _{\alpha }v(q_{\bar{q}})/k_{g,c\bar{c}}^{2}\cdot \{\T {n}{{}}{i}{c}{l}{{}}\T
 {j}{g}{l}{{}}{i}{{\bar{c}}}\T {n}{{}}{i}{q}{i}{{\bar{q}}}\},\\
 &  & T_{9}=e_q\cdot \bar{u}_{qg}\gamma ^{\alpha }v(q_{\bar{q}})\bar{u}(q_{c})\gamma
 _{\alpha }v_{\bar{c}\gamma }/k_{g,q\bar{q}}^{2}\cdot \{\T
 {n}{{}}{i}{c}{i}{{\bar{c}}}\T {j}{g}{i}{q}{l}{{}}\T
 {n}{{}}{l}{{}}{i}{{\bar{q}}}\},\\
 &  & T_{10}=e_q\cdot \bar{u}(q_{q})\gamma ^{\alpha }v_{\bar{q}g}\bar{u}(q_{c})\gamma
 _{\alpha }v_{\bar{c}\gamma }/k_{g,q\bar{q}}^{2}\cdot \{\T
 {n}{{}}{i}{c}{i}{{\bar{c}}}\T {n}{{}}{i}{q}{l}{{}}\T
 {j}{g}{l}{{}}{i}{{\bar{q}}}\},\\
 &  & T_{11}=e_q\cdot \bar{u}_{qg}\gamma ^{\alpha }v(q_{\bar{q}})\bar{u}_{c\gamma
 }\gamma _{\alpha }v(q_{\bar{c}})/k_{g,q\bar{q}}^{2}\cdot \{\T
 {n}{{}}{i}{c}{i}{{\bar{c}}}\T {j}{g}{i}{q}{l}{{}}\T
 {n}{{}}{l}{{}}{i}{{\bar{q}}}\},\\
 &  & T_{12}=e_q\cdot \bar{u}(q_{q})\gamma ^{\alpha }v_{\bar{q}g}\bar{u}_{c\gamma
 }\gamma _{\alpha }v(q_{\bar{c}})/k_{g,q\bar{q}}^{2}\cdot \{\T
 {n}{{}}{i}{c}{i}{{\bar{c}}}\T {n}{{}}{i}{q}{l}{{}}\T
 {j}{g}{l}{{}}{i}{{\bar{q}}}\},\\
 &  & T_{13}=e_q\cdot \bar{u}_{c,q\bar{q}}\hat{\epsilon }_{g}v_{\bar{c}\gamma }\cdot
 \{\T {n}{{}}{i}{c}{l}{{}}\T {j}{g}{l}{{}}{i}{{\bar{c}}}\T
 {n}{{}}{i}{q}{i}{{\bar{q}}}\},\\
 &  & T_{14}=e_q\cdot \bar{u}_{cg}\hat{J}_{q\bar{q}}v_{\bar{c}\gamma }\cdot \{\T
 {j}{g}{i}{c}{l}{{}}\T {n}{{}}{l}{{}}{i}{{\bar{c}}}\T
 {n}{{}}{i}{q}{i}{{\bar{q}}}\},\\
 &  & T_{15}=e_q\cdot \bar{u}_{cg}\hat{\epsilon }_{\gamma }v_{\bar{c},q\bar{q}}\cdot
 \{\T {j}{g}{i}{c}{l}{{}}\T {n}{{}}{l}{{}}{i}{{\bar{c}}}\T
 {n}{{}}{i}{q}{i}{{\bar{q}}}\},\\
 &  & T_{16}=e_c\cdot \bar{u}_{c,q\bar{q}}\hat{\epsilon }_{\gamma }v_{\bar{c}g}\cdot
 \{\T {n}{{}}{i}{c}{l}{{}}\T {j}{g}{l}{{}}{i}{{\bar{c}}}\T
 {n}{{}}{i}{q}{i}{{\bar{q}}}\},\\
 &  & T_{17}=e_q\cdot \bar{u}_{c\gamma }\hat{J}_{q\bar{q}}v_{\bar{c}g}\cdot \{\T
 {n}{{}}{i}{c}{l}{{}}\T {j}{g}{l}{{}}{i}{{\bar{c}}}\T
 {n}{{}}{i}{q}{i}{{\bar{q}}}\},\\
 &  & T_{18}=e_q\cdot \bar{u}_{c\gamma }\hat{\epsilon }_{g}v_{\bar{c},q\bar{q}}\cdot
 \{\T {j}{g}{i}{c}{l}{{}}\T {n}{{}}{l}{{}}{i}{{\bar{c}}}\T
 {n}{{}}{i}{q}{i}{{\bar{q}}}\},\\
 &  & T_{19}=e_c\cdot \bar{u}_{q,c\bar{c}}\hat{\epsilon }_{g}v_{\bar{q}\gamma }\cdot
 \{\T {n}{{}}{i}{q}{l}{{}}\T {j}{g}{l}{{}}{i}{{\bar{q}}}\T
 {n}{{}}{i}{c}{i}{{\bar{c}}}\},\\
 &  & T_{20}=e_c\cdot \bar{u}_{qg}\hat{J}_{c\bar{c}}v_{\bar{q}\gamma }\cdot \{\T
 {j}{g}{i}{q}{l}{{}}\T {n}{{}}{l}{{}}{i}{{\bar{q}}}\T
 {n}{{}}{i}{c}{i}{{\bar{c}}}\},\\
 &  & T_{21}=e_c\cdot \bar{u}_{qg}\hat{\epsilon }_{\gamma }v_{\bar{q},c\bar{c}}\cdot
 \{\T {j}{g}{i}{q}{l}{{}}\T {n}{{}}{l}{{}}{i}{{\bar{q}}}\T
 {n}{{}}{i}{c}{i}{{\bar{c}}}\},\\
 &  & T_{22}=e_q\cdot \bar{u}_{q,c\bar{c}}\hat{\epsilon }_{\gamma }v_{\bar{q}g}\cdot
 \{\T {n}{{}}{i}{q}{l}{{}}\T {j}{g}{l}{{}}{i}{{\bar{q}}}\T
 {n}{{}}{i}{c}{i}{{\bar{c}}}\},\\
 &  & T_{23}=e_c\cdot \bar{u}_{q\gamma }\hat{J}_{c\bar{c}}v_{\bar{q}g}\cdot \{\T
 {n}{{}}{i}{q}{l}{{}}\T {j}{g}{l}{{}}{i}{{\bar{q}}}\T
 {n}{{}}{i}{c}{i}{{\bar{c}}}\},\\
 &  & T_{24}=e_c\cdot \bar{u}_{q\gamma }\hat{\epsilon }_{g}v_{\bar{q},c\bar{c}}\cdot
 \{\T {j}{g}{i}{q}{l}{{}}\T {n}{{}}{l}{{}}{i}{{\bar{q}}}\T
 {n}{{}}{i}{c}{i}{{\bar{c}}}\},
\end{eqnarray}
The spinor states with the fixed spin projection on the axis \( z \)
have been chosen in the capacity of two independent spinor states.  In
our calculations, the Dirac representation for the \( \gamma \)-matrices has
been used. The spinor states can be written down as follows:
\begin{eqnarray}
  \displaystyle & & u(p,+)={\frac{1}{\sqrt{E+m}}\left(
      \begin{array}{c}
        E+m\\
        0\\
        p_{z}\\
        p_{x}+ip_{y}
\end{array}\right) ,}\nonumber \\
 &  & u(p,-)={\frac{1}{\sqrt{E+m}}\left( \begin{array}{c}
0\\
E+m\\
p_{x}-ip_{y}\\
-p_{z}
\end{array}\right) ,}\nonumber \\
 &  & v(p,+)={-\frac{1}{\sqrt{E+m}}\left( \begin{array}{c}
p_{z}\\
p_{x}+ip_{y}\\
0\\
E+m
\end{array}\right) ,}\nonumber \\
 &  & v(p,-)={\frac{1}{\sqrt{E+m}}\left( \begin{array}{c}
p_{x}-ip_{y}\\
p_{z}\\
0\\
E+m
\end{array}\right) .}
\end{eqnarray}
The \( \gamma \)-matrices in the Dirac representation are given by
\[
\gamma ^{0}=\left( \begin{array}{cc}
    1 & 0\\
    0 & -1
\end{array}\right) ,\qquad {\vec{\gamma }}=\left( \begin{array}{cc}
0 & {\vec{\sigma }}\\
-{\vec{\sigma }} & 0
\end{array}\right) ,\]

\[
{\vec{\sigma }}=\{\sigma _{x},\sigma _{y},\sigma _{z}\}=\left\{ \left(
    \begin{array}{cc}
      0 & 1\\
      1 & 0
\end{array}\right) ,\quad \left( \begin{array}{cc}
0 & -i\\
i & 0
\end{array}\right) ,\quad \left( \begin{array}{cc}
1 & 0\\
0 & -1
\end{array}\right) \right\} \]
The Gell-Mann matrices have the form
 \[
 \lambda ^{1}=\left( \begin{array}{ccc}
     0 & 1 & 0\\
     1 & 0 & 0\\
     0 & 0 & 0
\end{array}\right) ,\qquad \lambda ^{2}=\left( \begin{array}{ccc}
0 & -i & 0\\
i & 0 & 0\\
0 & 0 & 0
\end{array}\right) ,\qquad \lambda ^{3}=\left( \begin{array}{ccc}
1 & 0 & 0\\
0 & -1 & 0\\
0 & 0 & 0
\end{array}\right) ,\]

\[
\lambda ^{4}=\left( \begin{array}{ccc}
    0 & 0 & 1\\
    0 & 0 & 0\\
    1 & 0 & 0
\end{array}\right) ,\qquad \lambda ^{5}=\left( \begin{array}{ccc}
0 & 0 & -i\\
0 & 0 & 0\\
i & 0 & 0
\end{array}\right) ,\qquad \lambda ^{6}=\left( \begin{array}{ccc}
0 & 0 & 0\\
0 & 0 & 1\\
0 & 1 & 0
\end{array}\right) ,\]

\[
\lambda ^{7}=\left( \begin{array}{ccc}
    0 & 0 & 0\\
    0 & 0 & -i\\
    0 & i & 0
\end{array}\right) ,\qquad \lambda ^{8}=\frac{1}{\sqrt{3}}\left(
\begin{array}{ccc}
1 & 0 & 0\\
0 & 1 & 0\\
0 & 0 & -2
\end{array}\right) ,\]
and the \( t \)-matrices have been chosen as \( t^{i}=\frac{1}{2}\lambda ^{i}
\).

Let us remind of the antisymmetric constant \( f_{abc} \) values, which
have been required for our calculations
\[ \displaystyle
\begin{array}{c}
  f_{123}=1,\\
  f_{147}=-f_{156}=f_{246}=f_{257}=f_{345}=-f_{367}=\frac{1}{2},\\
  f_{458}=f_{678}=\frac{\sqrt{3}}{2}.
\end{array}\]

\subsection*{ B. The soft process of \(\boldsymbol c\)- and \( \boldsymbol
{\bar{q}}\)-quark fusion  into the \(\boldsymbol {(c\bar{q})}\)-quarkonium}

To describe the fusion of \( c \)- and \( \bar{q} \)-quarks into the 
\( (c\bar{q}) \)-quarkonium we suppose that there are terms in the
partonic distribution of \( D^{*} \)-meson, which correspond to the
valence quarks. We also suppose that the \( c \)- and \( \bar{q} \)-quarks
produced in the hard process transform into the valence quarks of the meson.
The valence quark distributions in the system of infinite momentum
have the following form:

\[
f_{c}^{v}(x,p_{\perp })=f_{c}(x,p_{\perp })-f_{\bar{c}}(x,p_{\perp
  }), \]
\[
f_{\bar{q}}^{v}(x,p_{\perp })=f_{\bar{q}}(x,p_{\perp
  })-f_{q}(x,p_{\perp }). \]
The averaged momentum fraction carried out by the valence quarks are
\begin{equation}
\label{valc}
\langle x_{c}^{v}\rangle =\int d^{2}p_{\perp }dxx\cdot f_{c}^{v}(x,p_{\perp
})\approx \frac{m_{c}}{m_{D^{*}}},
\end{equation}
\begin{equation}
\label{vald}
\langle x_{\bar{q}}^{v}\rangle =\int d^{2}p_{\perp }dxx\cdot
f_{\bar{q}}^{v}(x,p_{\perp })\approx \frac{\bar{\Lambda }}{m_{D^{*}}},
\end{equation}
where \( \bar{\Lambda } \) is the quark binding energy inside the
meson.

For \( \langle x^{v}_{c}\rangle \) and \( \langle
x^{v}_{\bar{q}}\rangle \) the following equation is valid
\begin{equation}
\label{valcd}
\langle x_{c}^{v}\rangle +\langle x_{\bar{q}}^{v}\rangle \approx 1 .
\end{equation}
In our calculation we neglect the dispersion of the momentum fractions
carried out by the quarks, or in other words, we consider
(\ref{valc},\ref{vald},\ref{valcd}) as absolutely precise equations.
Also we suppose that the light quark mass is about \( \bar{\Lambda }
\) .

In the doubly heavy quarkonium the quark pair is in the singlet state, because
the octet one is suppressed by the third power of the relative velocity of
quarks \cite{v3}. In the case of \( (c\bar{q}) \)-quarkonium this suppression
does not exist, and one has to take into account the both color states.

The operators \( \langle O_{(1)}\rangle \) and \( \langle
O_{(8)}\rangle \) describing the hadronization of singlet and octet
quark pairs into the meson are nonperturbative, because they include the
interaction on the scale about \( \Lambda _{\textrm{QCD}} \). In
the framework of nonrelativistic potential model these operators
correspond to the wave functions squared at the origin: \( \langle
O_{(1,8)}\rangle |_{\textrm{NR}}=|\Psi (0)|^{2} \).  They are defined
by
\begin{equation}
\label{O1}
\langle O_{(1)}\rangle =\frac{1}{12M_{D^{*}}}\left( -g^{\mu \nu }+\frac{p^{\mu
}p^{\nu }}{M^{2}_{D^{*}}}\right) \times \langle D^{*}(p)|(\bar{c}\gamma _{\mu
}q)(\bar{q}\gamma _{\nu }c)|D^{*}(p)\rangle ,
\end{equation}
\begin{equation}
\label{O8}
\langle O_{(8)}\rangle =\frac{1}{8M_{D^{*}}}\left( -g^{\mu \nu }+\frac{p^{\mu
}p^{\nu }}{M^{2}_{D^{*}}}\right) \times \langle D^{*}(p)|(\bar{c}\gamma _{\mu
}\lambda ^{a}q)(\bar{q}\gamma _{\nu }\lambda ^{b}c)|D^{*}(p)\rangle
\frac{\delta ^{ab}}{8}.
\end{equation}
The probability \( W(c\rightarrow D^{*}) \) of \( c \)-quark
fragmentation into the \( D^{*} \)-meson in \( e^{+}e^{-} \)-annihilation
can be expressed through the operators \( \langle O_{(1)}\rangle \) and \(
\langle O_{(8)}\rangle \) of our model
\begin{equation}
\label{W}
W(c\rightarrow D^{*})=\int _{0}^{1}D_{c\rightarrow D^{*}}(z)dz=\frac{\alpha
_{s}^{2}(\mu _{R})\langle O^{\textrm{eff}}(\mu _{R})\rangle }{m^{3}_{q}}I\left(
\frac{m_{q}}{m_{q}+m_{c}}\right) ,
\end{equation}
where \( \langle O^{\textrm{eff}}(\mu _{R})\rangle =\langle
O_{(1)}\rangle +\frac{1}{8}\langle O_{(8)}\rangle \), and
\begin{equation}
\label{I}
I(r)=\frac{8}{27}\left[
\frac{24+109r-126r^{2}-174r^{3}-89r^{4}}{15(1-r)^{5}}+\frac{r(7-4r-3r^{3}+10r^{
3}+2r^{4})}{(1-r)^{6}}\ln r\right] .
\end{equation}

\newpage
\includegraphics{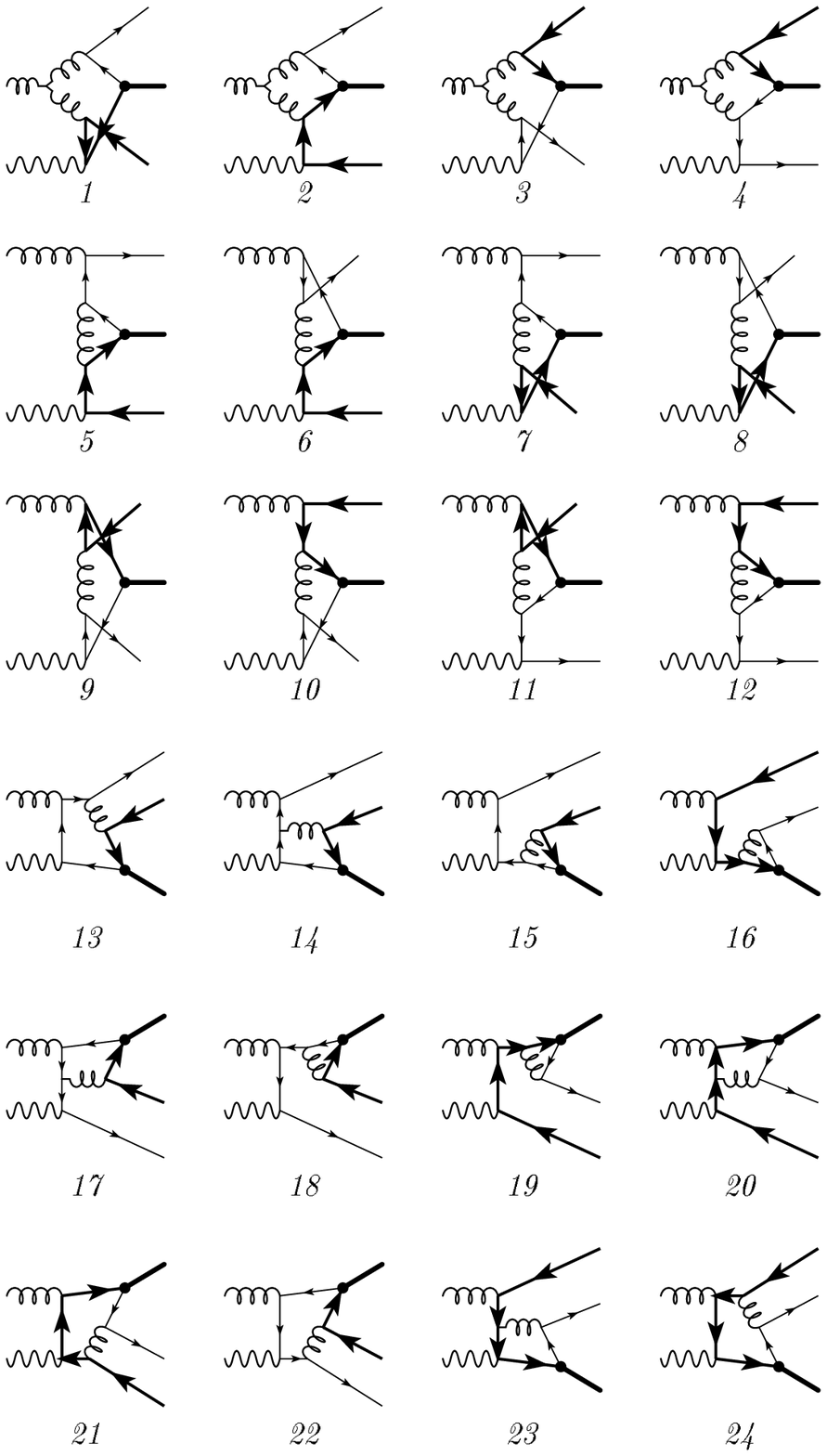} \vspace*{21
  cm}\hspace*{2 cm} Fig.~1. \parbox[t]{12cm}{The leading order QCD
  diagrams for the \( c\bar{q} \)-state production in the \( g\gamma ^{*}
  \)-interaction.} \newpage\vspace*{-3cm}\includegraphics{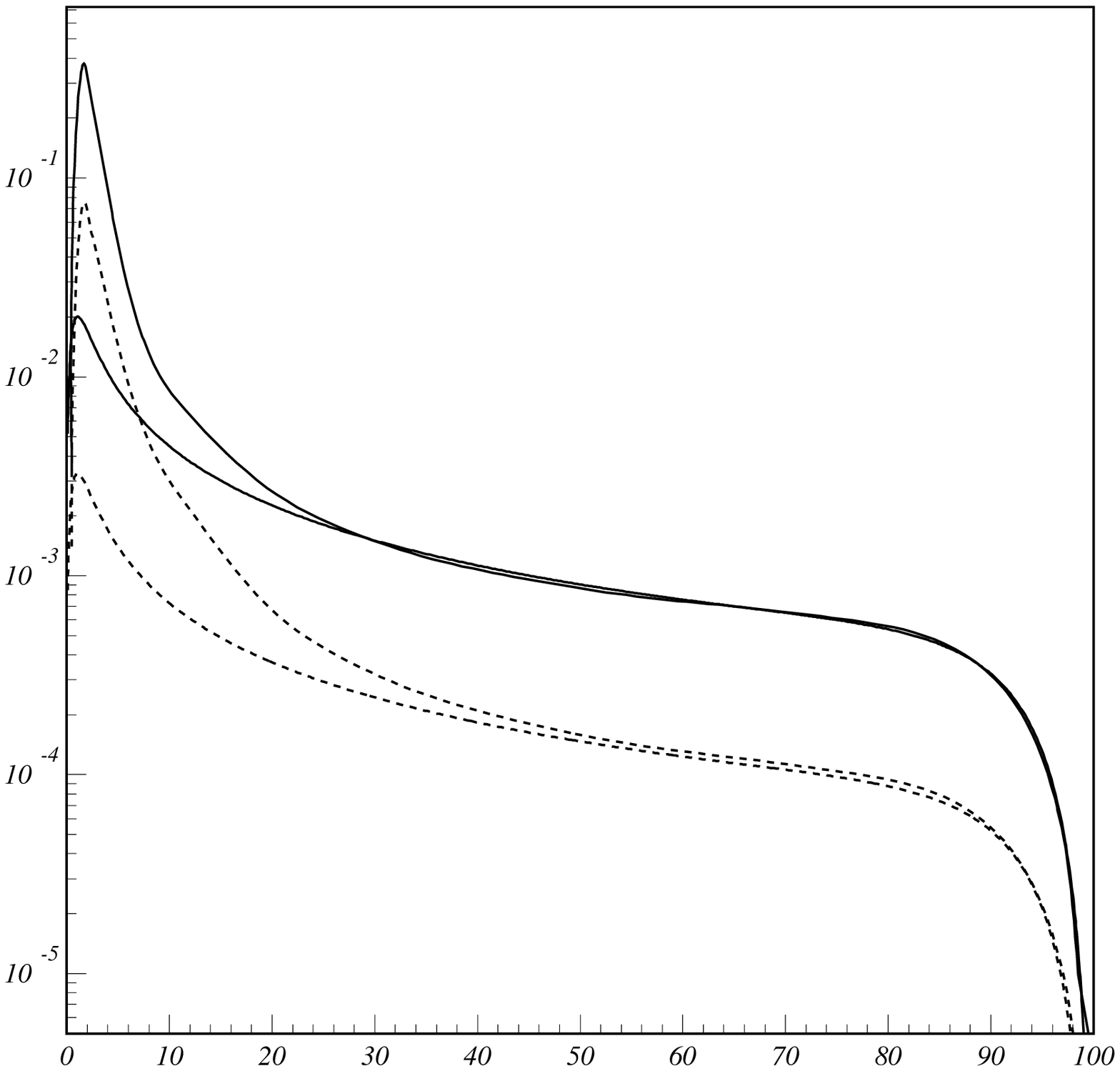}
\begin{picture}(450,450)\put(0,320){\( d\sigma _{g\gamma }/dp_{T} \) ,
    nb/GeV} \put(340,-50){\(p_{T}\), GeV} \put(20,-80){Fig.~2.
    \parbox[t]{12cm}{The \( D^{*} \)-meson production distribution versus \(
      p_{T} \) calculated in the framework of model under
      consideration (upper curve) in comparison with the fragmentation
      model prediction (lower curve) for the \( g\gamma \)-interaction at 200
      GeV.

}}
\end{picture} \newpage\vspace*{-3cm}\includegraphics{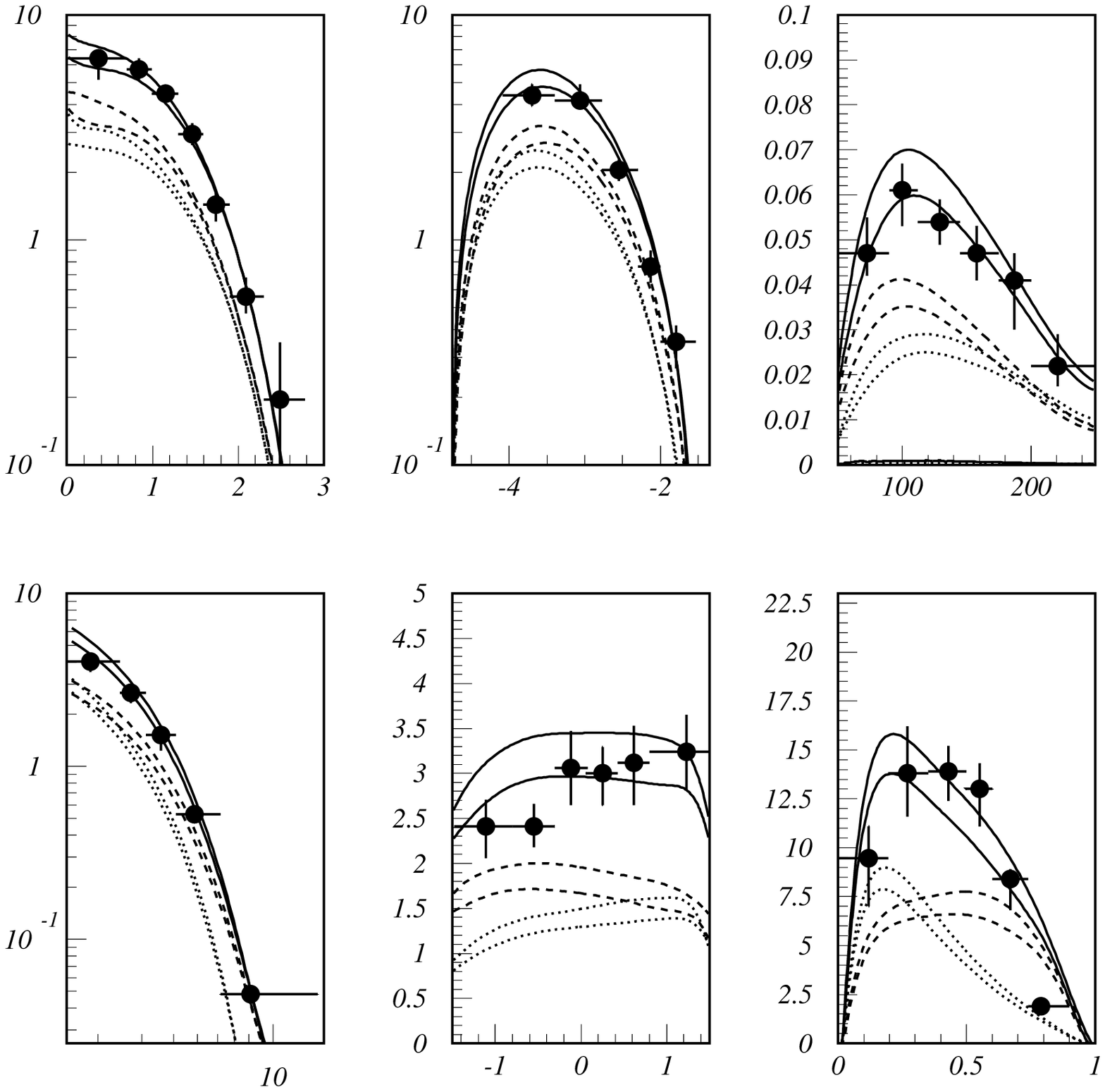} \begin{picture}(450,450)\put(-30,350){\(d\sigma/dlog_{10}Q^{2}\),
nb}
\put(140,350){\(d\sigma/dlog_{10}x\), nb}\put(305,350){\(d\sigma/dW\), nb/GeV}
\put(-30,100){\(d\sigma/dp_{T}\), nb/GeV}\put(140,100){\(d\sigma/d\eta\), nb}
\put(305,100){\(d\sigma/dx(D^{*})\), nb}\put(70,120){\(log_{10}Q^{2}\)}
\put(240,120){\(log_{10}x\)}\put(405,120){\(W\), GeV}\put(70,-130){\(p_{T}\),
GeV}
\put(240,-130){\(\eta(D^{*})\)}\put(405,-130){\(x(D^{*})\)}
\put(70,300){\large \textit{a}}\put(240,300){\large \textit{b}}
\put(405,300){\large \textit{c}}\put(70,50){\large \textit{d}}
\put(240,50){\large \textit{e}}\put(405,50){\large \textit{f}}
\put(20,-180){Fig.~3. \parbox[t]{12cm}
{The differential production cross sections of the \( D^{*} \)-meson
for the deep inelastic \( ep \)-scattering: over the photon virtuality
GeV\( ^{2} \)(\emph{a}), the Bjorken \( x \) (\emph{b}), the invariant mass of
the final hadrons (\emph{c}), the transverse momentum (\emph{d}), the
pseudorapidity (\emph{e}) and the Feynman variable (\emph{f})
in comparison with the experimental data of ZEUS Collaboration.}}

\end{picture}

\newpage \vspace*{-3 cm} \includegraphics{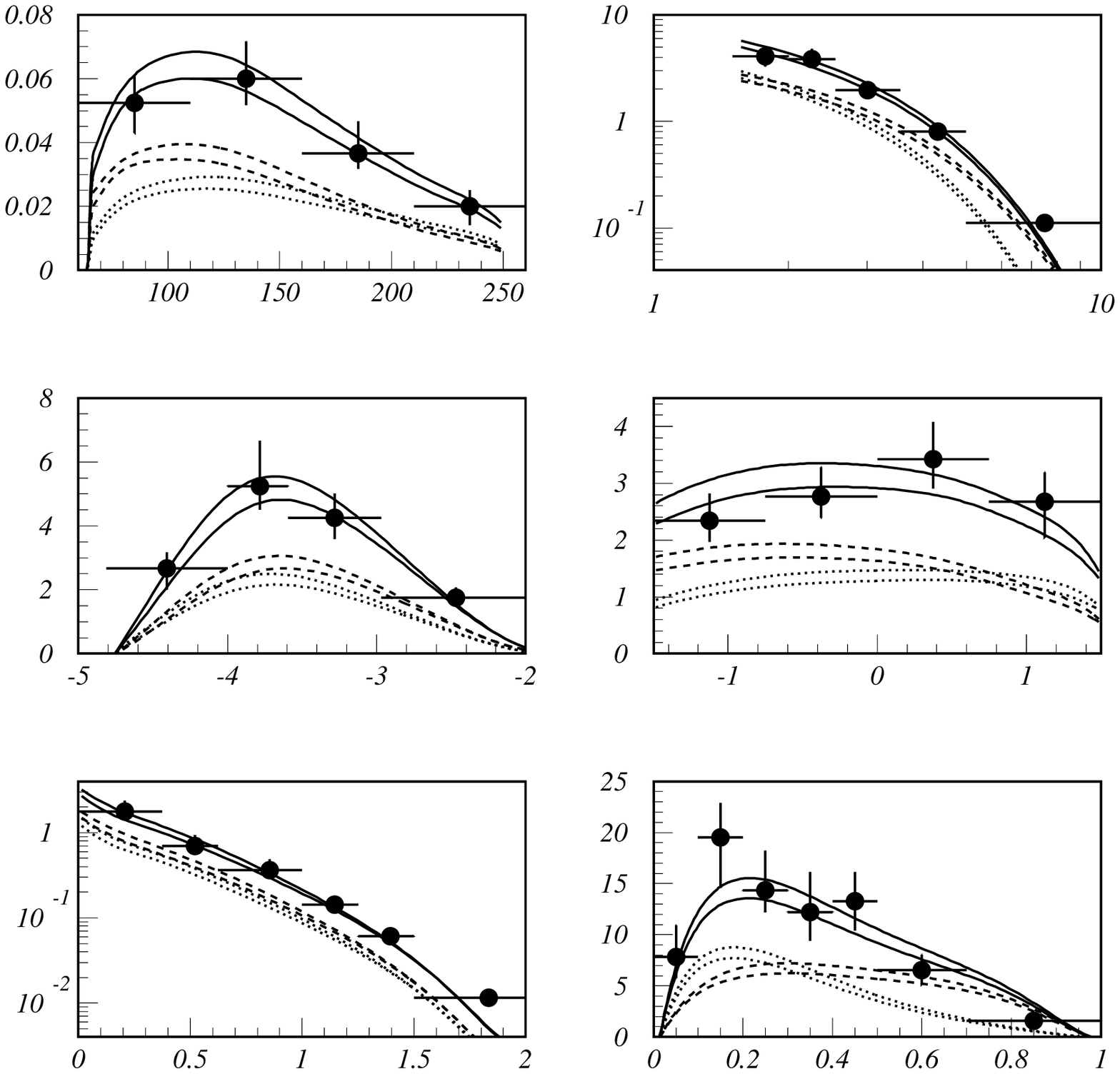}

\begin{picture}(450,450)
\put(240,15){\(d\sigma/dz\), nb}
\put(240,185){\(d\sigma/d\eta\), nb}
\put(240,350){\(d\sigma/dp_{T}\), nb/GeV}
\put(-10,15){\(d\sigma/dQ^{2}\), nb/GeV\(^{2}\)}
\put(-10,185){\(d\sigma/d\log_{10} x\), nb}
\put(-10,350){\(d\sigma/dW\), nb/GeV}
\put(400,-130){\(z(D^{*})\)}
\put(400,40){\(\eta(D^{*})\)}
\put(400,205){\(p_{T}\), GeV}
\put(150,-130){\(\log_{10} Q^{2}\)}
\put(150,40){\(\log_{10} x\)}
\put(150,205){\(W\), GeV}
\put(400,-10){\large \textit{f}}
\put(400,155){\large \textit{d}}
\put(400,325){\large \textit{b}}
\put(150,-10){\large \textit{e}}
\put(150,155){\large \textit{c}}
\put(150,325){\large \textit{a}}
\put(20,-180){Fig.~4 \parbox[t]{12cm}
{The differential production cross sections of the \( D^{*} \)-meson
for the deep inelastic \( ep \)-scattering over the invariant mass of
the final hadrons (\emph{a}), the transverse momentum (\emph{b}), the 
Bjorken \( x \) (\emph{c}), the pseudorapidity (\emph{d}), the photon
virtuality GeV\( ^{2} \)(\emph{e}) and \( z(D^{*}) \) in comparison 
with the experimental data of H1 Collaboration.}}
\end{picture}

\newpage\vspace*{-3cm}\includegraphics{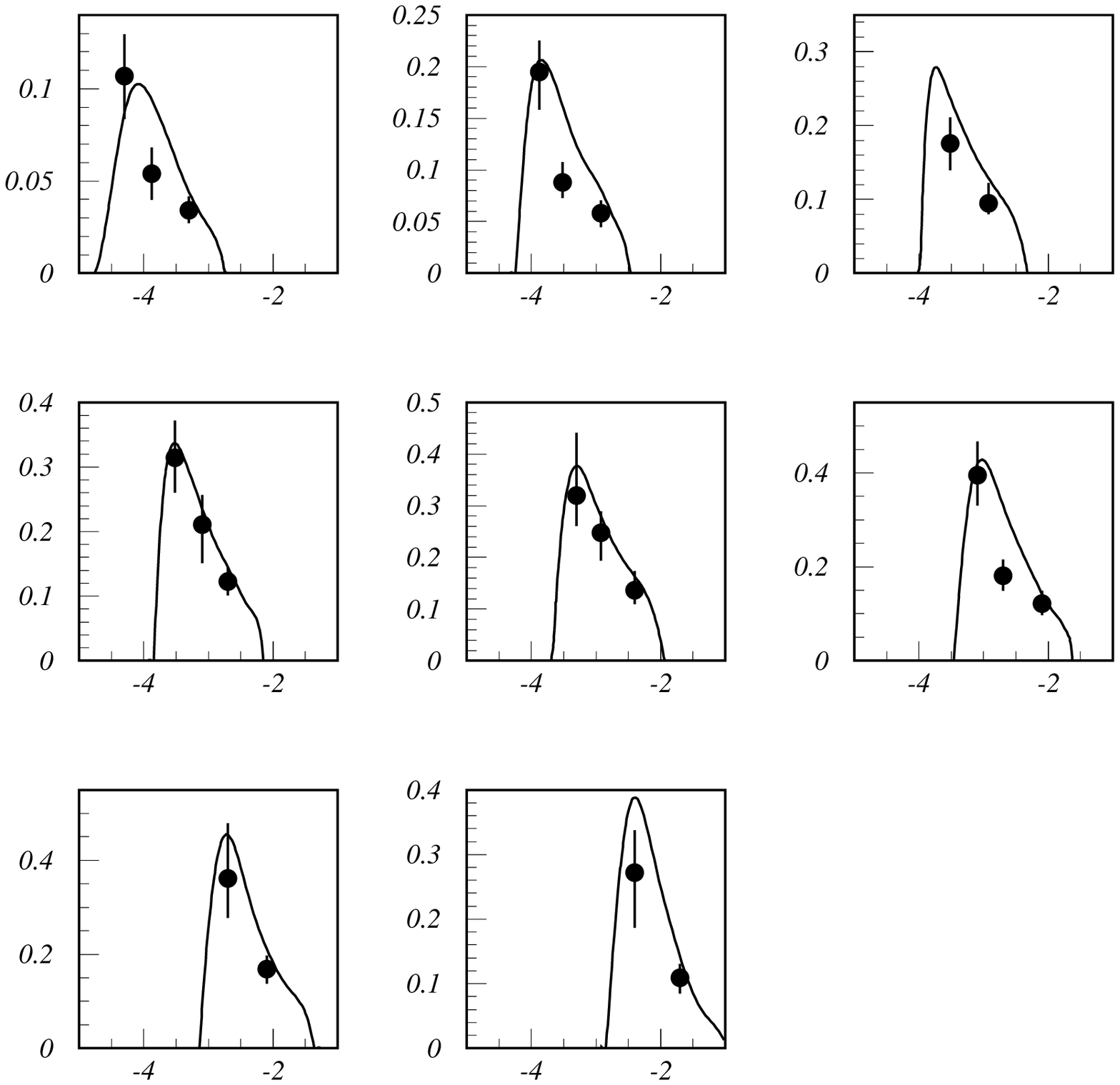}

\begin{picture}(450,450)
\put(-30,350){\(F_{2}^{c}\)}
\put(40,300){\(Q^{2}=1.8\)}
\put(50, 285){GeV\(^{2}\)}
\put(235,300){4}
\put(400,300){7}
\put(70,130){11}
\put(240,130){18}
\put(405,130){30}
\put(70,-30){60}
\put(240,-30){130}
\put(20,-180){Fig.~5. \parbox[t]{12cm}
{The experimental values of \( F^{c}_{2} \) (the ZEUS Collaboration)
in comparison with the discussed model predictions.}}
\end{picture}

\newpage\vspace*{-3cm}\includegraphics{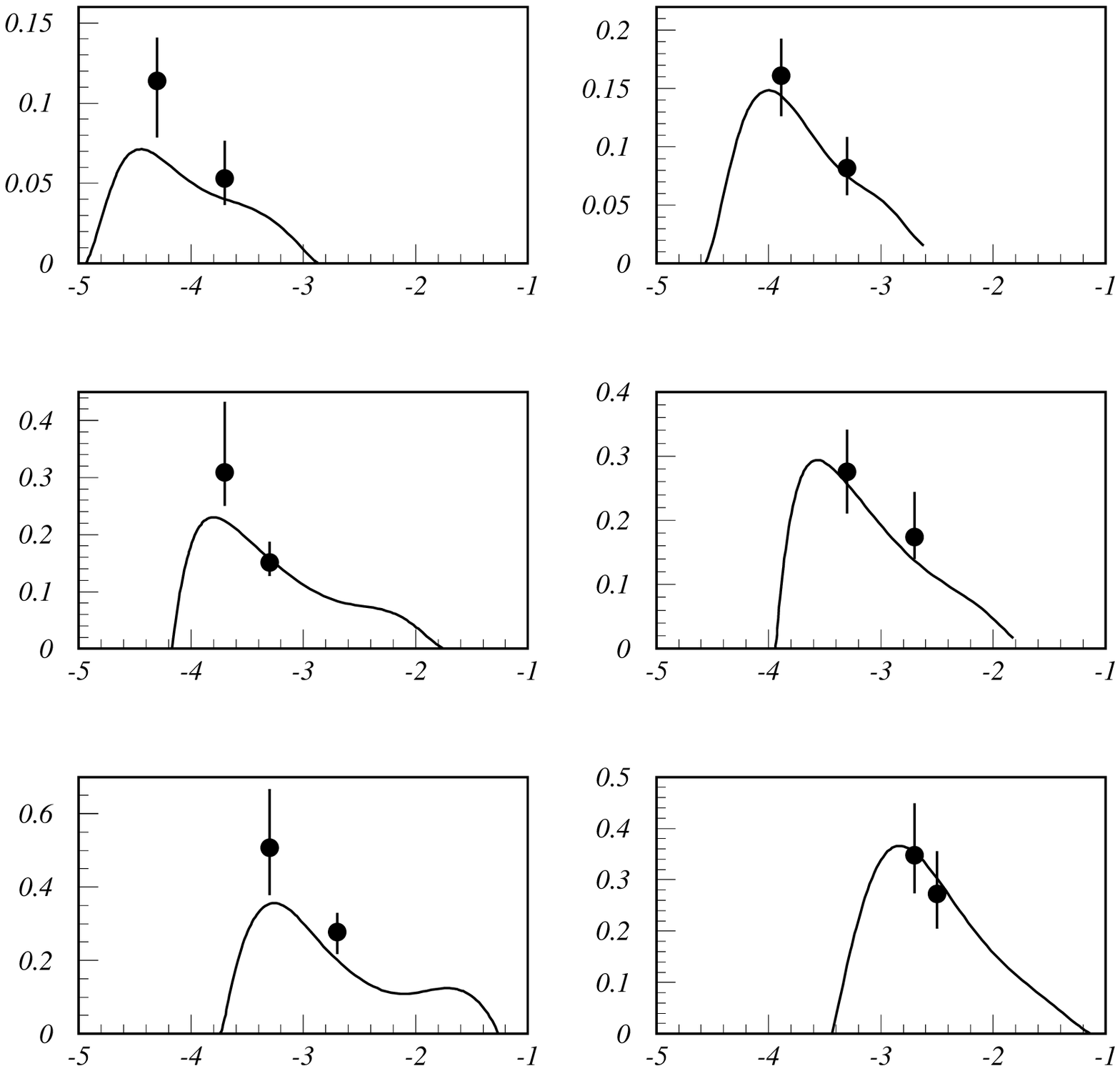}

\begin{picture}(450,450) 
\put(-10,350){\(F_{2}^{c}\)}
\put(400,-130){\(x\)} 
\put(400,-10){\(60\)}
\put(400,155){\(12\)}
\put(400,325){\(3.5\)}
\put(150,-10){\(25\)}
\put(150,155){\(6.5\)}
\put(100,325){\(Q^{2}=1.5\)GeV\(^{2}\)}

\put(20,-180){Fig.~6. \parbox[t]{12cm}
{The experimental values of \( F^{c}_{2} \) (the H1 Collaboration) in
comparison with the discussed model predictions.}}
\end{picture}

\newpage\includegraphics{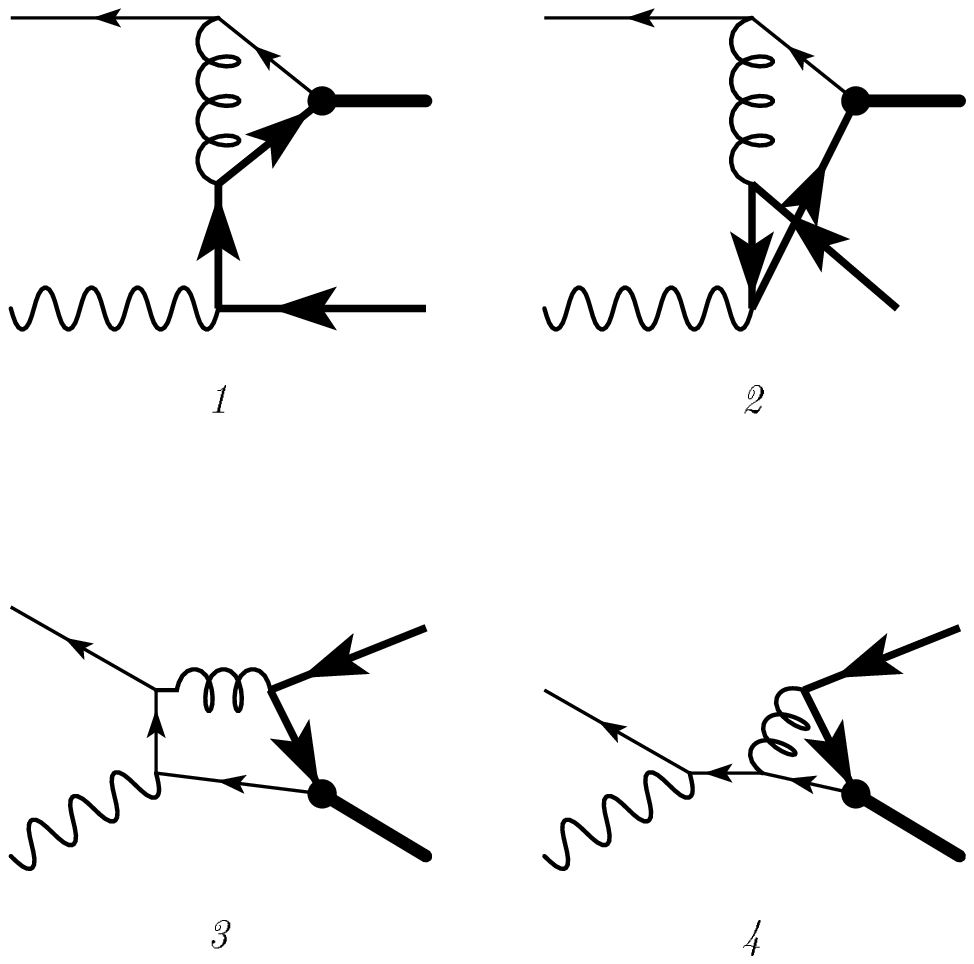} \vspace*{16
  cm} \hspace*{2 cm} Fig.~7. \parbox[t]{9cm}{The QCD leading order
  diagrams for the \( c\bar{q} \)-state production in the \( \bar{q}\gamma
  ^{*} \)-interaction.}

\newpage \vspace*{-3 cm} \includegraphics{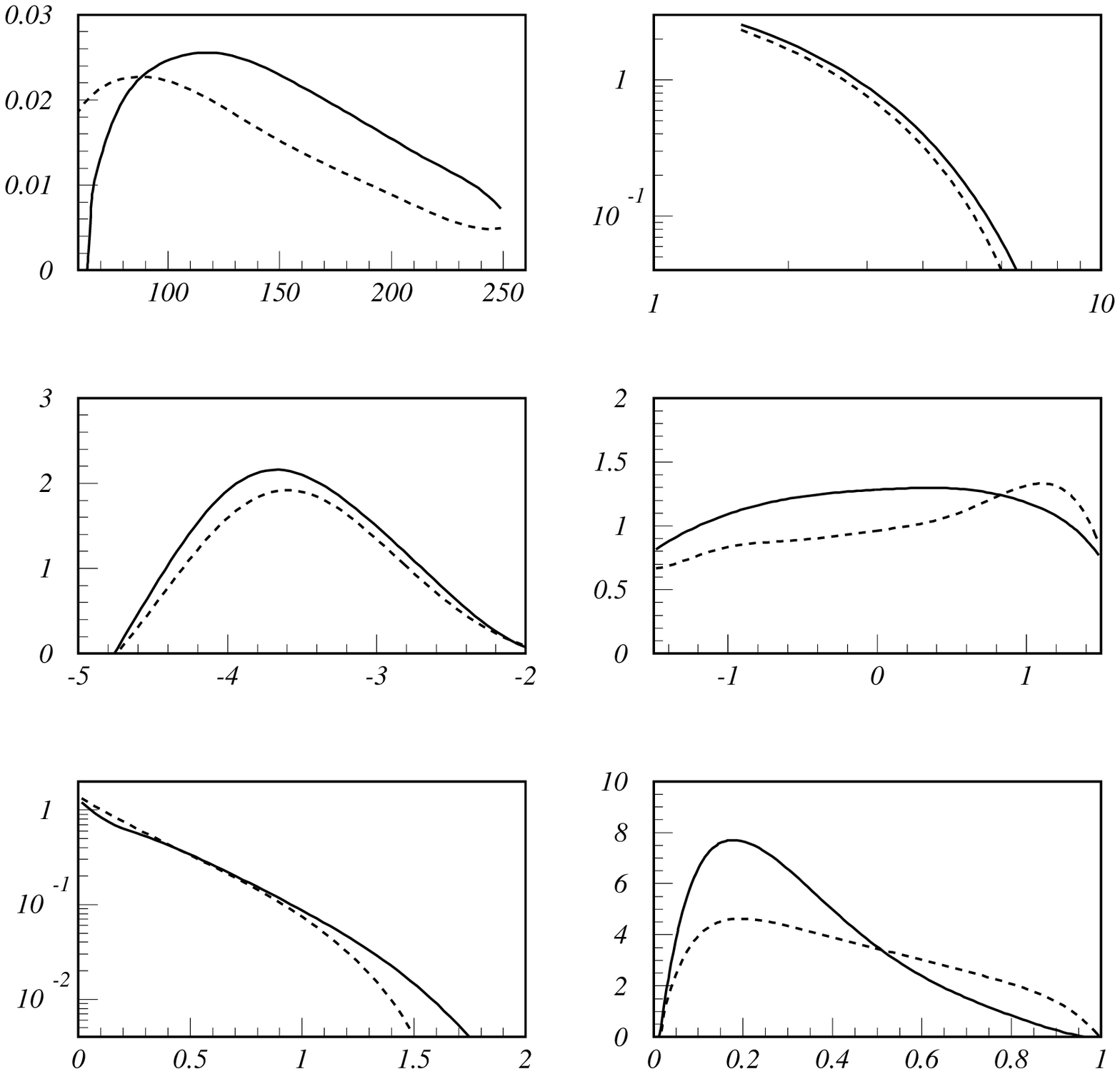}

\begin{picture}(450,450)
\put(240,15){\(d\sigma/dz\), nb}
\put(240,185){\(d\sigma/d\eta\), nb}
\put(240,350){\(d\sigma/dp_{T}\), nb/GeV}
\put(-10,15){\(d\sigma/dQ^{2}\), nb/GeV\(^{2}\)}
\put(-10,185){\(d\sigma/d\log_{10} x\), nb}
\put(-10,350){\(d\sigma/dW\), nb/GeV}
\put(400,-130){\(z(D^{*})\)}
\put(400,40){\(\eta(D^{*})\)}
\put(400,205){\(p_{T}\), GeV}
\put(150,-130){\(\log_{10} Q^{2}\)}
\put(150,40){\(\log_{10} x\)}
\put(150,205){\(W\), GeV}
\put(400,-10){\large \textit{f}}
\put(400,155){\large \textit{d}}
\put(400,325){\large \textit{b}}
\put(150,-10){\large \textit{e}}
\put(150,155){\large \textit{c}}
\put(150,325){\large \textit{a}}
\put(20,-180){Fig.~8 \parbox[t]{12cm}
{The singlet \( c\bar{q} \)-pair production in the \( \bar{q}\gamma -
\)subprocess for the kinematic region of H1 Collaboration in comparison with
the octet \( c\bar{q} \)-pair production in the \( g\gamma  \)-subprocess
(the distributions are over the same variables as in Fig.~4). }}
\end{picture}

\newpage \vspace*{-3 cm} \includegraphics{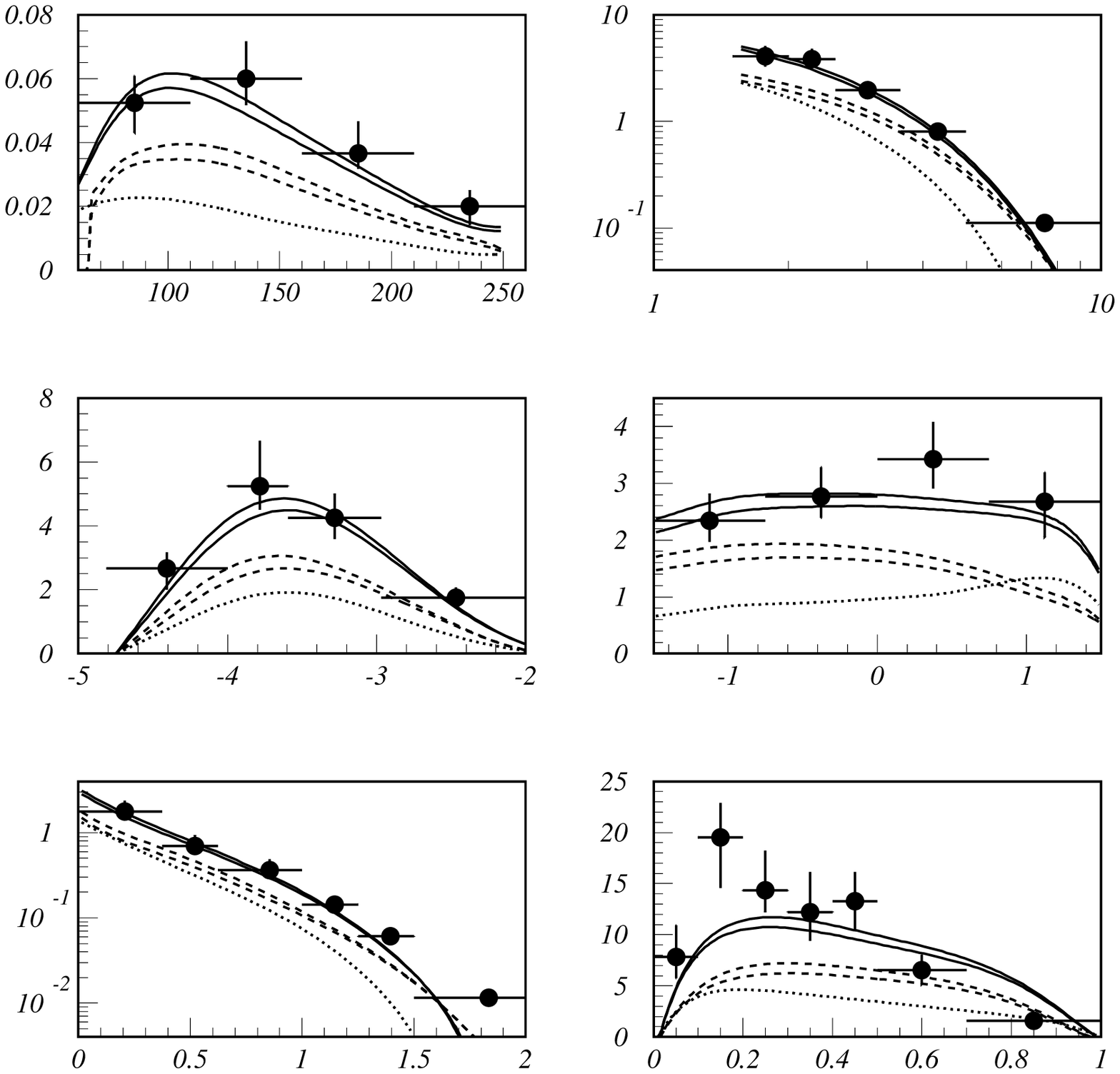}

\begin{picture}(450,450)
\put(240,15){\(d\sigma/dz\), nb}
\put(240,185){\(d\sigma/d\eta\), nb}
\put(240,350){\(d\sigma/dp_{T}\), nb/GeV}
\put(-10,15){\(d\sigma/dQ^{2}\), nb/GeV\(^{2}\)}
\put(-10,185){\(d\sigma/d\log_{10} x\), nb}
\put(-10,350){\(d\sigma/dW\), nb/GeV}
\put(400,-130){\(z(D^{*})\)}
\put(400,40){\(\eta(D^{*})\)}
\put(400,205){\(p_{T}\), GeV}
\put(150,-130){\(\log_{10} Q^{2}\)}
\put(150,40){\(\log_{10} x\)}
\put(150,205){\(W\), GeV}
\put(400,-10){\large \textit{f}}
\put(400,155){\large \textit{d}}
\put(400,325){\large \textit{b}}
\put(150,-10){\large \textit{e}}
\put(150,155){\large \textit{c}}
\put(150,325){\large \textit{a}}
\put(20,-180){Fig.~9 \parbox[t]{12cm}
{The sum of singlet \( c\bar{q} \)-pair production in the \( \bar{q}\gamma -
\)subprocess and the singlet \( c\bar{q} \)-pair production in the \( g\gamma
\)-subprocess in comparison with the data of H1 Collaboration (the
distributions are over the same variables as in Fig.~4).}}
\end{picture}
\end{document}